\documentclass[12pt]{iopart}
\usepackage{bm}
\usepackage[dvips]{graphicx}
\usepackage{color}
\usepackage{slashed}
\newcommand{\be}{\begin{equation}}

\newcommand{\ee}{\end{equation}}
\newcommand{\bea}{\begin{eqnarray}}
\newcommand{\eea}{\end{eqnarray}}

\newcommand{\dl}{\sum \!\!\!\!\!\!\!\! { \int}\,\, d \ell\,}
\begin{document}
\vspace*{-3cm}\hspace*{13cm} ICCUB-11-003\\  \hspace*{13.8cm} YITP-11-5\\  \hspace*{13.9cm}PHY-12921-TH-2011\\ \newline
\title{Collective modes in the color flavor locked phase}
\author{Roberto~Anglani$^1$, Massimo~Mannarelli$^2$ and  Marco~Ruggieri$^3$}
\address{$^1$Physics Division, Argonne National Laboratory, Argonne, Illinois 60439, USA\\
Institute of Intelligent Systems for Automation, CNR, I-70126 Bari, Italy}
\address{$^2$Departament d'Estructura i Constituents de la Mat\`eria and
Institut de Ci\`encies del Cosmos, Universitat de Barcelona,
Mart\'i i Franqu\`es 1, 08028 Barcelona, Spain\\ I.N.F.N., Laboratori Nazionali del Gran Sasso, Assergi (AQ), Italy}
\address{$^3$Yukawa Institute for Theoretical Physics, Kyoto University, 606-8502 Kyoto, Japan}
\date{\today}
\begin{abstract}
We  study the low energy effective action for some collective modes of the
color flavor locked  phase of QCD. 
This phase of matter  has long been known to be a superfluid because by picking a phase its order parameter breaks the
quark-number $U(1)_B$ symmetry spontaneously. 
We consider the  modes describing fluctuations in the magnitude
of the condensate, namely the Higgs mode, and in the phase of the condensate, namely the  Nambu-Goldstone (or Anderson-Bogoliubov) mode
associated with the breaking of $U(1)_B$. 
By employing  as microscopic theory the Nambu-Jona Lasinio model, we reproduce known results for
the  Lagrangian of the Nambu-Goldstone field to the leading order in the chemical potential and extend such results 
evaluating corrections due to the gap parameter.
Moreover, we determine the interaction terms between the Higgs and the
Nambu-Goldstone field. This study paves the way for a more reliable study of various
dissipative processes in rotating compact stars with a quark matter core in the
color flavor locked phase. 
\end{abstract}
 \maketitle

\section{Introduction}
At extremely high densities, Quantum chromodynamics (QCD) predicts that
the individual nucleons that form standard hadronic matter should melt and their  quark matter
content should be liberated~\cite{Collins}.
At  the  low temperatures expected in sufficiently old compact stars, 
 quark matter  is likely to be in one of the
possible color superconducting phases, whose critical temperatures
are generically of order tens of MeV~\cite{reviews,Nardulli:2002review}.
Since compact star temperatures are well below these
critical temperatures, for many purposes  the quark matter that may be
found within compact stars can be approximated as having zero temperature,
as we shall assume throughout. At asymptotic densities, where
the up, down and strange quarks can be treated on an equal footing
and  effects due to the strange quark mass can be neglected,
quark matter is in the color flavor locked (CFL)
phase~\cite{Alford:1998mk,reviews}. The CFL condensate is
antisymmetric in color and flavor indices and  involves
pairing between up, down and strange quarks. 
The order parameter  breaks the quark-number $U(1)_B$
symmetry spontaneously, and the
corresponding Nambu-Goldstone (NG) boson (Anderson-Bogoliubov
mode) determines the superfluid properties of  CFL
quark matter. 
The gapless excitation corresponds to a phase oscillation  about
the mean field value of the gap parameter. The fluctuation in magnitude of the
condensate is associated with a massive mode, which we shall refer to as the
Higgs mode. Both of these fluctuations are excitations of several fermions and therefore describe collective modes of the system. 

The low energy properties of the system are completely determined by these collective modes~\cite{Casalbuoni:1999wu}, and their study is mandatory
to understand dynamical properties that take place  in compact stars with a CFL core. The actual superfluid property of the system is
due to the presence of the gapless excitations, which satisfy the Landau's criterion for superfluidity~\cite{superfluids}. 
Moreover, in rotating superfluids the interactions between NG bosons and  vortices  lead to the appearance of the so-called mutual friction force between the normal and the superfluid components of the system. The NG bosons are also responsible of many other properties  of cold superfluid
matter, in particular they contribute to the thermal conductivity and to the shear and bulk
viscosities~\cite{superfluids}.  A  study of the shear viscosity and of the standard bulk viscosity coefficient  due to NG bosons has been done in Refs.~\cite{Manuel:2004iv,Manuel:2007pz}. A more detailed study of the bulk viscosity
coefficients of the CFL phase has been done in~\cite{Mannarelli:2009ia}. In the CFL phase also the global $SU(3)_R \times SU(3)_L$ symmetry is spontaneously broken and the corresponding pseudo-NG bosons can contribute to the transport properties of the CFL phase.  The contribution of kaons to the bulk and shear viscosity  has  been studied in Refs.~\cite{Alford:2007rw,Alford:2009jm}. The contribution of NG bosons to the thermal conductivity and cooling of compact stars were studied in Refs.~\cite{Shovkovy:2002kv,Braby:2009dw, Reddy:2002xc, Jaikumar:2002vg}.
Other low energy degrees of freedom are the plasmons, which have been studied in Refs.~\cite{Casalbuoni:2000na, Gusynin:2001tt}. 

Although the CFL phase is characterized by many low energy degrees of freedom, in the present paper we shall focus on the Nambu-Goldstone field, $\phi$, and on the Higgs field, $\rho$, because we aim to build the  effective Lagrangian for the description of NG boson-vortex interaction. Vortices are described by spatial variation of the condensate and therefore are determined by the space variation of the Higgs field. In principle one could study interaction of vortices with other low energy degrees of freedom, but the NG boson associated to the braking of $U(1)_B$ is the only massless mode, whereas pseudo-NG bosons associated to the breaking of chiral symmetry have a mass of the order of few keV~\cite{Son:1999cm}, and plasmons associated to the breaking of $SU(3)_c$ symmetry have even larger  masses~\cite{Casalbuoni:2000na, Gusynin:2001tt}. Therefore all of these modes are   thermally suppressed in cold compact stars.

We derive the  effective Lagrangian of the Nambu-Goldstone field,  and of the Higgs field  using as  microscopic theory the Nambu-Jona Lasinio model~\cite{Nambu:1961tp, revNJL} with a local four-Fermi interaction with the quantum numbers of one gluon exchange. This model mimics some aspects of QCD at large densities~\cite{Buballa:2003qv, Ebert:2006tc}. We derive the interaction terms of the NG bosons up to terms of the type $(\partial \phi)^4$. The leading contributions to these interaction terms were obtained by Son in Ref.~\cite{Son:2002zn}  using symmetry arguments and the expression of the
pressure of the CFL phase. To our knowledge the results of Ref.~\cite{Son:2002zn} so far have not been obtained starting from a microscopic theory. Only the free Lagrangian was obtained in~\cite{Son:1999cm}.  We extend the results of~\cite{Son:2002zn} including the next to leading corrections proportional to the gap parameter. In order to do this we  integrate out the fermionic degrees of freedom employing the High Density Effective Theory (HDET)~\cite{Nardulli:2002review}. Moreover we determine the kinetic Lagrangian for the Higgs field and the interaction terms between the Higgs and the NG bosons. 

A similar analysis to the one presented here was done for non-relativistic
systems at unitarity in~\cite{Schakel:2009} and in~\cite{Gubankova:2008ya, Escobedo:2010uv}. As we shall discuss in
Section~\ref{sec-higgs}, the main difference with the non-relativistic systems at unitarity is  that in the CFL phase
integrating out the Higgs field does not lead to a change of the speed of sound.

This paper is organized as follows. In Section~\ref{sec-model} we present the
Nambu-Jona Lasinio model and we determine the expression of the effective action of the
system in the HDET approximation. In Section~\ref{sec-phonon} we derive the
effective Lagrangian for the NG boson, neglecting the oscillations in the
magnitude of the condensate.  In Section~\ref{sec-higgs} we derive the LO interaction terms between the
Higgs mode and the NG bosons and the kinetic Lagrangian for the Higgs field.
We also compare our results with the non-relativistic results of
\cite{Schakel:2009}.
We draw our conclusions in Section~\ref{sec-conclusion}. Some interaction vertices as well as the calculation of some integrals  are reported in the Appendix.

\section{The model}\label{sec-model}
We consider a Nambu-Jona Lasinio (NJL) model of quark matter with a local four-Fermi interaction as a  model of
QCD at large quark chemical potential $\mu$. We  assume that  $\mu \gg m_s$ and therefore we neglect $u$, $d$
and $s$ quark masses.
In the absence of interactions the Lagrangian density describing the system of up, down and strange quarks  is
given by
\begin{equation}
{\cal L}_0=\bar\psi_{i\alpha}(i\slashed{\partial} +\mu
\gamma_{0})\psi_{i\alpha} \, ,
\end{equation} 
where $i,j=1,2,3$ are flavor indices, $\alpha,\beta=1,2,3$ are
color indices and the Dirac indices have been suppressed. 
In QCD one can show that at large densities the gluon exchange leads to the formation of a quark-quark
condensate~\cite{reviews}. At
asymptotic densities the favored phase is the color flavor locked phase~\cite{Alford:1998mk} which is characterized by
the condensate
\be
\langle\psi^t_{i\alpha}C\psi_{j\beta}\rangle\sim\Delta \sum_{I=1,2,3}\epsilon_{\alpha\beta I} \epsilon^{i j I}\,,
\ee
which locks together color and flavor rotations. Here  $C= i \gamma^2\gamma^0$ is  the charge conjugation matrix
and $\epsilon_{\alpha\beta I}$ and $\epsilon^{i j I}$ are Levi-Civita tensors.
This condensate breaks several symmetries of QCD, see~\cite{Alford:1998mk}  for a detailed analysis. For our purposes it
is sufficient to note that the CFL condensate breaks the $U(1)_B$ symmetry and therefore
leads to the appearance of a gapless Nambu-Goldstone boson. 

In the NJL model the  interaction among quarks mediated by
gluons is replaced by a   four-Fermi interaction of the BCS type.
We shall consider an interaction with the same quantum numbers of one gluon exchange 
\begin{equation}
{\cal L}_{\rm
I}=-\frac{3 }{16} g\, \bar \psi \gamma_\mu\lambda^{A}\psi \bar \psi\gamma^\mu \lambda^{A}\psi\,,
\end{equation}
where $g>0$ is the coupling constant  and $\lambda^A$ with $A=1,...,8$ are the Gell-Mann matrices.  This interaction at
large chemical potentials  leads to the formation of the CFL condensate. Introducing  the new basis for the quark fields
\be
\psi_{i \alpha} = \frac{1}{\sqrt{2}} \sum_{A=1}^9 \lambda^A_{i
\alpha}\psi_A \,,
\ee
where  $\lambda^9 =  \sqrt{2/3}\times I$, we have that
 \begin{equation}
\langle\psi^t_{A}C\psi_{B}\rangle\sim\Delta_{AB}\,,
\end{equation}
where  $\Delta_{AB}=\Delta_{A} \delta_{AB}$, where $\Delta_1=...=\Delta_8=\Delta$ and $\Delta_9 = - 2
\Delta$. 
 In this basis  the four-Fermi interaction is given by
\begin{equation}
{\cal L}_{\rm
I}=-\frac{g}{4}V_{ABCD}\epsilon_{ab}\epsilon_{\dot{c}\dot{d}}\psi^{A}_{a}\psi^{B}_{b}\psi^{C\dag}_{\dot{c}}\psi^{D\dag}_
{\dot{d}}\,,
\end{equation}
where $\displaystyle V_{ABCD}={\rm Tr}\sum_{E=1}^8(\lambda_A\lambda_E\lambda_B\lambda_C\lambda_E\lambda_D) $ and $a(\dot{a})=1,2$ are the Weyl indices for $L(R)$ components~\cite{Casalbuoni-review}.

In order to study the fluctuation of the condensate we introduce the Hubbard-Stratonovich fields $\Delta_{AB}(x)$
and $\Delta^{*}_{AB}(x)$ which allow to write the partition
function (normalized at the free case for $g=0$) as
\be
\frac{{\cal Z}}{{\cal Z}_{0}}=
\frac{1}{{\cal Z}_{0}}\int[d\psi,d\psi^{\dag}][d\Delta,d\Delta^{*}]\exp\left\{i\int d^{4}x\;\left[
-\frac{\Delta_{AB}W_{ABCD}\Delta^{*}_{CD}}{g}+{\cal L}_{\Delta}\right]\right\}\,,
\ee
where
\be
W_{ABCD}V_{CDEF} =\delta_{AE}\delta_{BF}  \qquad {\rm and} \qquad V_{ABCD}W_{CDEF}=\delta_{AE}\delta_{BF}\,,
\ee
and the semi-bosonized Lagrangian is given by
\begin{equation}
{\cal L}_{\Delta}=\bar\psi_A(i\slashed{\partial}+\mu\gamma_{0})\psi_A-\frac{1}{2}\Delta_{AB}(\psi_A^\dag
C\psi_B^*)+\frac{1}{2}\Delta^{*}_{AB}(\psi_{A}^tC\psi_{B})\,.
\end{equation}

The fluctuations of the condensate
$\Delta(x)$ around the mean field value $\Delta^{MF}$ can be described by two real fields; one  field describes 
the variation of $|\Delta(x)|$ while the other  field is associated to a local phase change.
Therefore we write
\begin{equation}\label{def-delta}
\Delta_{AB}(x)=[\Delta_{AB}^{MF}+\rho_{AB}(x)]e^{2i\phi(x)}\,,
\end{equation}
where $\rho_{AB}(x)=\rho(x)\, {\rm diag}(1,...,1,-2)_{AB}$ and $\phi(x)$ are the  real fields.
Hereafter we shall suppress the indices $A,B$ and we shall indicate with $\Delta$ the
mean field value of the gap in order to simplify the notation.   
We find   convenient to redefine the fermionic fields as
\begin{equation}
\psi \to \psi\, e^{i \phi(x)} \,,
\end{equation}
and in this  way the semi-bosonized Lagrangian is given by
\bea\label{expansion}
{\cal L}_{\Delta} &=& \bar \psi \left(i \gamma^\mu \partial_{\mu}+
\gamma^0\mu -\gamma^{0}\partial_{0}\phi-\gamma^i \partial_i  \phi \right) \psi \nonumber\\ &-&\frac{1}{2} \psi^{\dagger}
C(\Delta +
\rho)\psi^* +\frac{1}{2} \psi^{t} C(\Delta + \rho)  \psi\,. \eea

Writing the Lagrangian  in this form it is possible to define an
``effective chemical potential'' 
\be\label{effectivemu} 
\tilde \mu = \mu - \partial_0 \phi\,.
\ee 
In other words $\partial_0 \phi$ describes long-wavelength
fluctuations of the chemical potential on the top of its constant value $\mu$. We want now to clarify one point
regarding the terminology used for the field $\phi$. Since  $\partial_0 \phi$ describes fluctuations of the chemical
potential, it is not correct to call this field the {\it phonon}, which describes pressure oscillations. However, when
one neglects the effect of the gap $\Delta$, the Lagrangian of the phonon and of the NG boson associated with the
breaking of $U(1)_B$ symmetry coincide. The reason is that for vanishing values of $\Delta$ the oscillations of pressure
are proportional to the oscillations of the chemical potential.

It is convenient to define a fictitious gauge field $A^{\mu} = (\partial_0 \phi,\nabla
\phi)$ which according with Eq.~(\ref{expansion}) is minimally coupled to the quark fields. 
Hereafter we shall refer to $A^{\mu}$
as the gauge field, although it is not related to any gauge symmetry of the system.
With this substitution  we rewrite the Lagrangian in Eq.~(\ref{expansion}) as 
\begin{equation}\label{expansion2}
{\cal L}_{\Delta} = \bar \psi \left(i \gamma^\mu D_{\mu}+
\mu\gamma^0  \right) \psi - \frac{1}{2} \psi^{\dagger} C(\Delta +
\rho)\psi^* +\frac{1}{2} \psi^{t}C(\Delta + \rho)  \psi\,, \end{equation}
where $D_{\mu}=\partial_{\mu}+iA_{\mu}$.

In order to simplify the calculation we employ the High Density Effective
Theory (HDET), see ~\cite{Nardulli:2002review}. Using standard techniques, the Lagrangian describing 
the kinetic terms and the interaction with the gauge field $A^\mu$ can be written as
\bea\label{HDET1}
{\cal L}_{I}&=&\int\frac{d{\bf v}}{8\pi}\left[\psi^{\dag}_{+}\left(iV\cdot
D-\frac{P^{\mu\nu}D_{\mu}D_{\nu}}{2\mu+i\tilde V\cdot D}\right)\psi_{+}\right.\nonumber\\ &+&\left.
\psi^{\dag}_{-}\left(i \tilde V\cdot
D-\frac{P^{\mu\nu}D_{\mu}D_{\nu}}{2\mu+i V\cdot D}\right)\psi_{-}\right]\,,
\eea
where the positive energy fields with ``positive" and ``negative" velocities are given by
\be \psi_{\pm}\equiv\psi_{+}(\pm v)\,,\ee 
and where
\be
V^{\mu}=(1,{\bf v} )\,, \qquad \tilde V^{\mu}=(1,-{\bf v})\,,  \qquad P^{\mu\nu} = g^{\mu\nu} -\frac{V^\mu \tilde V^\nu
+V^\nu \tilde V^\mu}{2}\,,
\ee
with ${\bf v}$ the Fermi velocity. 
The non-local interactions which are present in Eq.~(\ref{HDET1}) are due to the integration of the negative energy
fields~\cite{Nardulli:2002review}.

In order to have a compact notation it is useful to introduce the Nambu--Gorkov spinor
\begin{equation}
\Psi=\frac{1}{\sqrt{2}}\left(\begin{array}{c}\psi_{+}
\\C\psi_{-}^{*}\end{array}\right) \,,
\end{equation}
which allows to write the semi-bosonized Lagrangian as
\begin{equation}\label{full-lagrangian}
{\cal L}_{\Delta}={\cal L}_{a}+{\cal L}_{b}\,,
\end{equation}
where 
\begin{equation}
{\cal L}_{a}=\int\frac{d{\bf
v}}{4\pi}\Psi^{\dag}\left(\begin{array}{cc}iV^\mu \partial_\mu-V^{\mu}
A_{\mu}
& \Delta+\rho \\  \Delta+\rho & i\tilde{V}^\mu \partial_\mu+\tilde V^{\mu}
A_{\mu}\end{array}\right)\Psi\,,
\end{equation}
is the sum of the kinetic term and  of the local interaction term,
while the non-local interaction term is given by
\begin{equation}\label{nonlocal}
{\cal L}_{b}=-\int\frac{d{\bf v}}{4\pi}P^{\mu\nu}\Psi^{\dag}\left(\begin{array}{cc}\frac{-2\mu+iV\cdot
D^*}{L}D_\mu D_\nu & \frac{\Delta}{L}D_\mu^* D_\nu  \\ \frac{\Delta}{L}D_\mu D_\nu^* &
\frac{2\mu + i \tilde V\cdot D}{L}D_\mu^*
D_\nu^*\end{array}\right)\Psi\,,
\end{equation}
where 
 $L=(2 \mu + i \tilde V \cdot
D)(-2 \mu + i V \cdot D^*) - \Delta^2 - i \epsilon$. More details about the derivation of the non-local interaction will
be given in~\cite{Mannarelli:future}. 

The semi-bosonized Lagrangian  is quadratic in
the fermionic fields and therefore we can integrate them out.  In this way  the effective action can be written in terms
of the fields $\phi$ and $\rho$ and
their derivatives. An analogous calculation in the non-relativistic case has been done in~\cite{Gubankova:2008ya}. 
Before doing this we consider in more detail the interaction terms between the fermionic fields and the fictitious
gauge field determined by the non-local term. In momentum space  we have that
\begin{equation}\label{Lb}
{\cal L}_b=\int\frac{d{\bf v}}{4\pi}\Psi^{\dag}P^{\mu\nu}A_\mu A_\nu \left(\begin{array}{cc} -2 \mu + V\cdot \ell +
V \cdot A & -\Delta \\-\Delta & 2 \mu + \tilde V\cdot \ell -
 \tilde  V \cdot A
\end{array}\right)\frac{1}{L}\Psi\,,
\end{equation}
where in momentum space $L=(2 \mu + \tilde V \cdot
\ell - \tilde V \cdot A )(-2 \mu +  V \cdot \ell + V \cdot A) - \Delta^2 - i \epsilon$.
The ``residual momentum" of the quarks,  $\ell^\mu$, is defined as follows:
\be 
\ell_0 = p_0 \qquad {\ell_i} = {p_i} - \mu {v_i}\,,  
\ee
where $p^\mu$ is the four-momentum of the quarks.  

\begin{figure}
\centering
\includegraphics[scale=0.5]{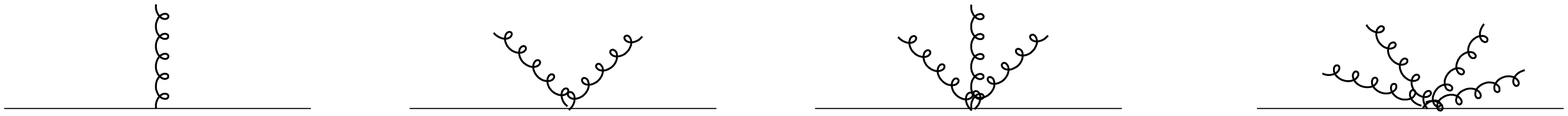}
\caption{The tree-level interaction vertices of the  gauge field with the fermions in the HDET. Full lines correspond
to  fermionic fields;  wavy lines correspond to the gauge fields. The vertices are named $\Gamma_1$, $\Gamma_2$,
$\Gamma_3$ and $\Gamma_4$, where the subscripts indicates the number of the external gauge fields. Their expression is
reported in the~\ref{appendix-vertices}.}
\label{fig-gamma-gauge}
\end{figure}

Expanding the denominator of the expression in Eq.~(\ref{Lb}) in powers of $A$ and considering terms up to the order
$A^4$, we have that
\bea
{\cal L}_{b}&=&\int\frac{d{\bf v}}{4\pi}\left( \Psi^{\dag}\Gamma_2^{\mu\nu}\Psi A_\mu A_\nu +
\Psi^{\dag}\Gamma_3^{\mu\nu\rho}\Psi A_\mu A_\nu A_{\rho} \right.\nonumber\\ &+& \left.
\Psi^{\dag}\Gamma_4^{\mu\nu\rho\sigma}\Psi A_\mu A_\nu A_{\rho} A_{\sigma} \right) + {\cal O}(A^5)\,,
\eea
where the expression of the vertices $\Gamma_2$, $\Gamma_3$ and $\Gamma_4$  are reported in the~\ref{appendix-vertices}.
We also define 
\be
\Gamma_{1}^{\mu}=\left(\begin{array}{cc} -V^\mu & 0 \\0 & \tilde V^\mu\end{array}\right) \,,
\ee
which is the vertex that describes the minimal coupling of quarks with the gauge field in the HDET. The
various vertices are schematically depicted in Fig.~\ref{fig-gamma-gauge}, with the wavy lines corresponding to the
gauge field and the full line corresponding to the fermionic field. 

Integrating out the fermionic fields, the partition function turns out to be given by
\be
\frac{{\cal Z}}{{\cal Z}_{0}}=
\frac{\int[d\Delta,d\Delta^{*}]\exp\left[\frac{1}{g}\int
d^{4}x\;\Delta_{AB}W_{ABCD}\Delta^{*}_{CD}\right]\det[S^{-1}]^{1/2}}{\det[S_{0}^{-1}]^{1/2}}\equiv\exp\left[i{\cal S}
\right]
\ee
where ${\cal S}$ is the  action of the system and the full inverse propagator is given by
\be
S^{-1} \equiv S^{-1}_{MF}+\Gamma\,.
\ee
The mean field inverse propagator is given by
\be
S^{-1}_{MF}=\left(\begin{array}{cc}i V^\mu \partial_\mu & -\Delta
\\-\Delta & i \tilde{V}^\mu \partial_\mu\end{array}\right)  \,,
\ee
while $\Gamma =\Gamma_\rho + \Gamma_{1}+\Gamma_{2} + \Gamma_{3}+\Gamma_{4}$. 
The interaction of quarks with the $\rho$ field is described by the vertex
\be\label{gammarho}
\Gamma_{\rho}=\left(\begin{array}{cc} 0 & -\rho \\-\rho & 0\end{array}\right)\,,  
\ee
while the interaction of quarks with the NG bosons is given by $\Gamma_{1}= \Gamma_{1}^{\mu} A_\mu$,
$\Gamma_{2}= \Gamma_{2}^{\mu\nu} A_\mu A_\nu$, $\Gamma_{3}= \Gamma_{3}^{\mu\nu\rho} A_\mu A_\nu A_\rho$ 
and $\Gamma_{4}= \Gamma_{4}^{\mu\nu\rho\sigma} A_\mu A_\nu A_\rho A_\sigma$.

\begin{figure}
\centering
\includegraphics[scale=0.5]{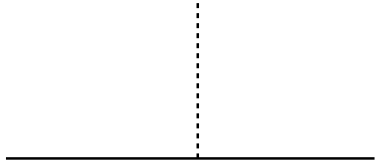}
\caption{Interaction of the Higgs field (dashed line) with  quarks (full line). The corresponding vertex is named
$\Gamma_\rho$ and is reported in Eq.~(\ref{gammarho}).}
\label{fig-higgs}
\end{figure}

We separate the mean field action from the fluctuation writing 
\begin{equation}
{\cal S}={\cal S}_{MF}+{\cal S}_{\rm eff}\,,
\end{equation}
where ${\cal S}_{\rm eff}$ is the effective action describing the low energy properties of the system. The mean
field action provides the free energy of the system
\begin{equation}
\Omega={\cal S}_{ MF}=-\frac{i}{2}{\rm
Tr}\ln[S^{-1}_{MF}]-\frac{i}{g}\left[\Delta_{AB}W_{ABCD}\Delta^{*}_{CD}\right] \,,
\end{equation}
and is a function of  the quark gap parameter $\Delta$. Hereafter, ${\rm
Tr}$, symbolizes the trace over the Nambu-Gorkov index, the trace over color-flavor  indices, the trace over spinorial indices and the trace over a complete set of functions in space-time.

The gap parameter can be determined by the stationary condition of the mean field action 
\be
\frac{\partial{\cal S}_{MF}}{\partial  \Delta}\Big|_{\bar\Delta} = 0 \,,
\ee
which in turn allows to determine the pressure  of the system by $P=-\Omega(\bar\Delta)$. 
In the NJL model, the mean field value of the gap parameter turns out to depend on the coupling $g$ and on the
three-momentum cutoff, see {\it e.g.}~\cite{Nardulli:2002review}. Since we do not need the numerical value of the gap, 
we shall treat $\Delta$ as a free parameter. 

The fluctuation around the mean field solution are described by the effective action
\bea\label{effective1}
{\cal S}_{\rm eff}&=&-\frac{i}{g}\int
d^{4}x\left[\rho_{AB}(x)W_{ABCD}\rho_{CD}(x)+2\rho_{AB}(x)W_{ABCD}\Delta_{CD}\right] \nonumber\\ &-&\frac{i}{2}{\rm
Tr}\ln\left(1+S_{MF}\Gamma\right)\,,
\eea
which contains the $\Gamma$ expansion 
\be\label{effective2} 
{\rm Tr}\ln\left(1+S_{MF}\Gamma\right)={\rm
Tr}\left[\sum_{n=1}^{\infty}\frac{(-1)^{n+1}}{n}(S_{MF}\Gamma)^{n}\right]\,,
\ee
and we shall evaluate terms  up to the fourth order in the gauge fields. We shall also  determine the leading order
terms of the  $\rho$ field  Lagrangian and the leading order interaction terms of the $\rho$ field with the NG bosons.
In the expansion we shall neglect the NLO terms of the kind $(\partial \partial \phi)^n$ where $n$ is any nonzero
integer. 

\section{The Lagrangian for the NG boson}\label{sec-phonon}
Neglecting  oscillations in the modulus of the condensate we can determine 
from the Eqs.~(\ref{effective1}) and (\ref{effective2}) the effective Lagrangian for the NG bosons. 
As we shall show below, for vanishing values of $\Delta$ we obtain the same results obtained in Ref.~\cite{Son:2002zn} 
\be\label{full-phi}
{\cal L}_{\phi} =  \frac{3}{4
\pi^2} \left[(\mu - \partial_0 \phi)^2 -(\partial_{i} \phi)^{2}\right]^2 \,.
\ee
This expression relies on  symmetry considerations, in particular on conformal symmetry, and on the expression of the
pressure in the CFL phase.
We shall reproduce these results and evaluate the corrections of the order $(\Delta/\mu)^2$
which  are related with the breaking of conformal symmetry.
Our strategy is to  first expand the Lagrangian in the $A_\mu$ fields, and
write
\be\label{L-phonon}
{\cal L}_{\phi}= {\cal L}_1+{\cal L}_2+{\cal L}_3+{\cal L}_4 \,,
\ee
where ${\cal L}_m$ is the term with $m$ gauge fields. Then we expand the various terms in $\Delta/\mu$.
Notice that the term ${\cal L}_m$ will be obtained by the expansion in Eq.~(\ref{effective2}) considering all the terms
with $n \le m$.

A different way of obtaining the leading order  in $\mu$ Lagrangian for terms proportional to $\partial_0 \phi$ is the
following. At the leading order in $\mu$ --- neglecting terms proportional to $\Delta$ --- the free-energy density of
the CFL phase in the absence of oscillations is given by
\be \Omega = \frac{3}{4 \pi^2} \mu^4 \,.\ee
We have seen in Section~\ref{sec-model} that  $\partial_0 \phi$ corresponds to a fluctuation of the chemical potential
of the system. Therefore, including these fluctuations  the free-energy density of the CFL phase at the leading order
in $\mu$ is given by 
\be \tilde \Omega = \frac{3}{4 \pi^2} \tilde\mu^4 \,,\ee
where $ \tilde\mu$ is defined in Eq.~(\ref{effectivemu}).
 Expanding the free energy we obtain that the Lagrangian for the $\partial_0 \phi$ field is given by
\be\label{omega}
{\cal L}(\partial_0 \phi)  =
\frac{3}{4\pi^{2}}\mu^{4}-\frac{3}{\pi^{2}}\mu^{3}\partial_{0}\phi+\frac{9}{2\pi^{2}}\mu^{2}
(\partial_{0}\phi)^{2}-\frac{3\mu}{\pi^{2}}(\partial_{0}\phi)^{3}+\frac{3}{4\pi^{2}}(\partial_{0}\phi)^{4}\,.
\ee

\subsection{One-point function}
\begin{figure}
\centering
\includegraphics[scale=0.5]{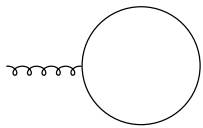}
\caption{One loop diagram that contributes to the coefficient of the  $\partial_0 \phi$ term in the effective Lagrangian
${\cal L}_1$ in Eq.~(\ref{L-phonon}). The full line corresponds to the quark field. The wavy line
corresponds to the external  gauge field.}
\label{fig-1points}
\end{figure}

The evaluation of the term of the effective Lagrangian proportional to $\partial \phi$ 
requires the  computation of the diagram in Fig.~\ref{fig-1points}. One has that
\be
{\cal L}_1 = -i  {\rm Tr}[S \Gamma]\Big|_{A} \,,
\ee
where the subscript means that in the evaluation of the trace  only the terms linear in $A$ must be included.
We find that
\be\label{L1}
{\cal L}_1  = -i \dl \left[\tilde V \cdot  A  \frac{V\cdot \ell
}{D^*} -  V \cdot A \frac{\tilde V\cdot \ell }{D}\right]\,,
\ee
where we have defined
\be  \dl = \frac{2}{\pi} \sum_{N_c,N_f}\int \frac{d {\bf v}}{4 \pi} \int_{-\delta}^{+\delta} \frac{d \ell_\parallel}{2
\pi} (\mu + \ell_{\parallel})^2 \int_{-\infty}^{\infty} \frac{d \ell_0}{2
\pi}\,, \ee
where $\delta$ is a cutoff that we shall set equal to $\mu$, see~\cite{Nardulli:2002review}. 
In Eq.~(\ref{L1}) we have also used~\cite{footnote}
\be
D = V \cdot \ell \tilde V \cdot \ell - \Delta^2 + i \epsilon \,,
\ee
and for future convenience we  define the quantity
\be\label{L0}
L_0 = (2 \mu + \tilde V \cdot \ell)(-2 \mu + V \cdot \ell) -\Delta^2 -i \epsilon \,.
\ee
The sum over flavor and color degrees of freedom is straightforward and we obtain 
\bea
{\cal L}_1 &=&  -\frac{2 i}{\pi} \int \frac{d {\bf v}}{4 \pi}\int \frac{d^2 \ell}{(2 \pi)^2}(\mu +
\ell_{\parallel})^2\left[\tilde
V \cdot  A \left(8\frac{V\cdot \ell }{D(\Delta)^*}+\frac{V\cdot \ell}{D(-2\Delta)^*}\right) \right.\nonumber\\  &-&
\left. V \cdot  A  \left(8\frac{\tilde V\cdot \ell }{D(\Delta)}+\frac{\tilde V\cdot \ell}{D(-2\Delta)}
\right) \right]\,. \eea
Employing the expressions reported in the~\ref{appendix-integrals} one can do the  integration over the
residual momentum and on the Fermi velocity. In this way one finds that  
\be
{\cal L}_1 = \left(-\frac{3}{\pi^2}\mu^3+\frac{6}{\pi^2}\mu^2\Delta\right) \partial_0 \phi  \,.
\ee
Notice that at the leading order in $\mu$ the expression above for the term proportional to $\partial_0 \phi$ agrees
with the corresponding expression reported in Eq.~(\ref{omega}).

\subsection{Two-point function}
\begin{figure}[htbp!]
\centering
\includegraphics[scale=0.5]{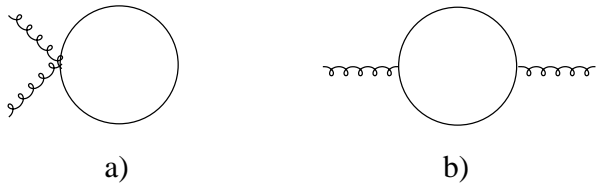}
\caption{One loop diagrams that contribute to the coefficients of the $A_0^2$ term, and of the ${\bf A}^2$ term of the
effective Lagrangian ${\cal L}_2$ in Eq.~(\ref{L-phonon}). Full
lines are fermionic fields. Wavy lines are gauge fields.}
\label{fig-2points}
\end{figure}

The Lagrangian involving two gauge fields is given by 
\be
{\cal L}_2 = -i\left. \left({\rm Tr} [S \Gamma] -\frac{1}2 {\rm Tr} [S \Gamma S \Gamma] \right)\right|_{A^2}\,,
\ee
where  in the evaluation of the trace one has to consider only terms quadratic in $A$. The corresponding diagrams are
reported in
Fig.~\ref{fig-2points}. The diagram in Fig.~\ref{fig-2points}a) gives
\be {\rm Tr}[S \Gamma]\Big|_{A^2} =  \dl  P^{\mu\nu}A_{\mu}A_\nu\left[  \frac{\Delta^2 + \tilde V\cdot \ell (V\cdot \ell -2 \mu)
}{L_0 D }  +  (V\to \tilde V)\right]
\,,
\ee
where $L_0$ has been defined in Eq.~(\ref{L0}) and $(V \to \tilde V)$ actually means an expression that is obtained by replacing $(V \to \tilde V,
\ell_{\parallel} \to -\ell_{\parallel}, \epsilon \to -\epsilon)$. Hereafter we shall always use this way of writing in
order to simplify the notation. 

\begin{figure}[htbp!]
\centering
\includegraphics[scale=0.4]{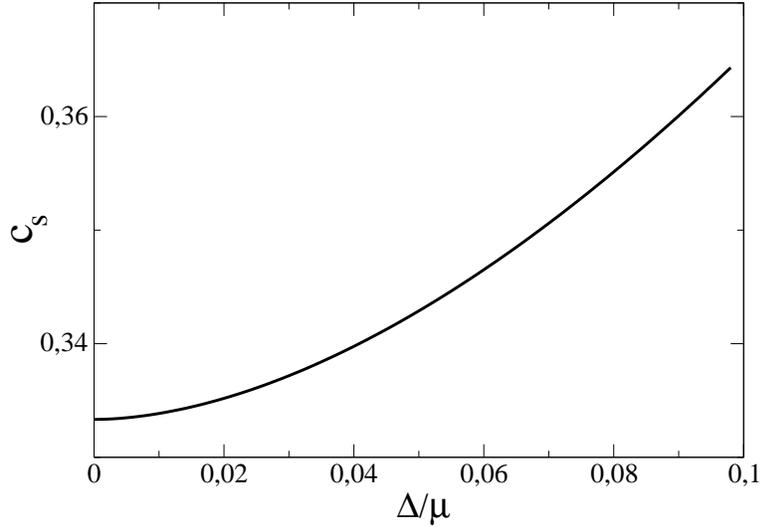}
\caption{Speed of the NG boson associated to the breaking of $U(1)_B$ as a function of $\Delta/\mu$.}
\label{fig-cs}
\end{figure}

The diagram in Fig.~\ref{fig-2points}b) gives
\be
 {\rm Tr}[S \Gamma S \Gamma]\Big|_{A^2} = \dl \left[\left(\frac{(\tilde V \cdot
A)^2(l_0+ \ell_\parallel)^2 }{D^2} - \frac{(\tilde V \cdot A  V \cdot A)\Delta^2 }{D^2}\right)+
(V \to \tilde V)\right] \,,
\ee
and evaluating the integrals using the expressions reported in the~\ref{appendix-integrals}  we have that
\bea
{\cal L}_2 &\simeq& \frac{9 \mu^2}{2 \pi^2}\left(1-2 \frac{\Delta^2}{\mu^2} \right) A_0^2  -
\frac{3 \mu^2}{2 \pi^2}\left(1-2.1 \frac{\Delta^2}{\mu^2} \right) {\bf A}^2 + {\cal
O}\left(\frac{\Delta^2}{\mu^2}\log{(\Delta/\mu)}\right)\nonumber\\&=& \frac{1}2 m_D^2
A_0^2 -  \frac{1}2 m_M^2 {\bf A}^2\,.
\label{2-point}
\eea
Then  the speed of the NG boson is given by $c_s=m_M/m_D$ and the corresponding plot is  reported in
Fig.~\ref{fig-cs}.  For vanishing $\Delta$ we have that $c_s = 1/\sqrt 3$ meaning that the system is scale invariant.
The effect of a nonvanishing $\Delta$ is to increase the speed of the NG boson.

\subsection{Three-point function}

\begin{figure}
\centering
\includegraphics[scale=0.5]{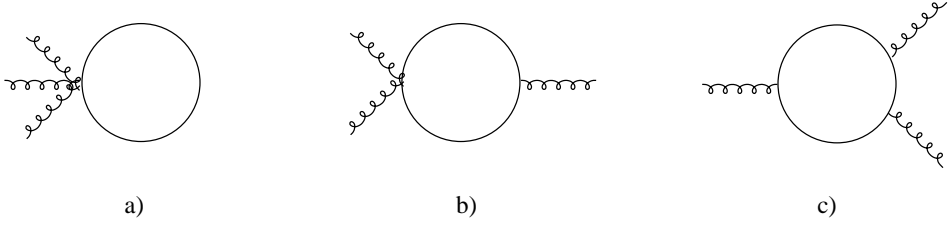}
\caption{One loop diagrams that contribute to the
effective Lagrangian ${\cal L}_3$ in Eq.~(\ref{L-phonon}). Notice that 
the only contribution to $A_0^3$ comes from diagram c). Full lines are fermionic fields. Wavy lines are gauge fields.}
\label{fig-3points}
\end{figure}

The Lagrangian describing the interaction of three NG bosons is formally given by
\be
{\cal L}_3 = -i\left. \left({\rm Tr} [S \Gamma] -\frac{1}2 {\rm Tr} [S \Gamma S \Gamma] +\frac{1}3 {\rm Tr} [S \Gamma S
\Gamma S \Gamma]\right)\right|_{A^3} \,,
\ee
and corresponds to the evaluation of the diagrams in  Fig.~\ref{fig-3points}.  The diagrams \ref{fig-3points}a) and
\ref{fig-3points}b) contribute exclusively to the term $A_0 {\bf A}^2$, while the diagram \ref{fig-3points}c) gives  the term proportional to $A_0^3$.  The contribution of the diagram in Fig.~\ref{fig-3points}a) is given by
\bea
 {\rm Tr} [S \Gamma]\Big|_{A^3} &=&\dl P^{\mu\nu}A_{\mu}A_\nu  V \cdot A \left[  \frac{L_0 \tilde V\cdot
\ell+(V\cdot\ell-\tilde V \cdot \ell - 4 \mu)(D+2\Delta^2-2\mu \tilde V\cdot\ell)}{D  L_0^2}\right.\nonumber\\ &-&
\left. (V \to \tilde V)\right]\,.
\eea

The contribution of the diagram in Fig.~\ref{fig-3points}b) is given by
\bea
{\rm Tr} [S \Gamma S \Gamma]\Big|_{A^3} &=& -2 \dl P^{\mu\nu}A_{\mu}A_\nu V \cdot A \left[ \frac{2\Delta^2 \tilde V
\cdot \ell+(V\cdot \ell -2 \mu)((\tilde V \cdot \ell)^2 -\Delta^2)}{D^2 L_0} \right.\nonumber\\ &-&
\left.(V \to \tilde V)\right] \,.
\eea

The contribution of the diagram in Fig.~\ref{fig-3points}c) is given by
\bea
{\rm Tr} [S \Gamma S \Gamma S \Gamma]\Big|_{A^3} &=& \dl\left[ \frac{- (V \cdot A)^3 (\tilde V\cdot \ell)^3-
\Delta^2(\tilde V \cdot A)^2 V \cdot A V \cdot \ell}{D^3} \right. \nonumber\\  
&+&\left.\frac{2 \Delta^2(\tilde V \cdot A)^2(V \cdot A)
(\tilde V\cdot \ell)}{D^3}
 -  (V \to \tilde V   )\right] \,.
\eea
Evaluating the integrals with the help of the expressions reported in the
\ref{appendix-integrals}  one finds that
\be
{\cal L}_3 = \frac{3 \mu}{\pi^2}\left(1-\frac{\Delta^2}{\mu^2}\right)A_0 {\bf A}^2-\frac{3
\mu}{\pi^2}\left(1-\frac{3\Delta^2}{2\mu^2}\right)A_0^3\,.
\ee

\subsection{Four-point function}
\begin{figure}
\centering
\includegraphics[scale=0.5]{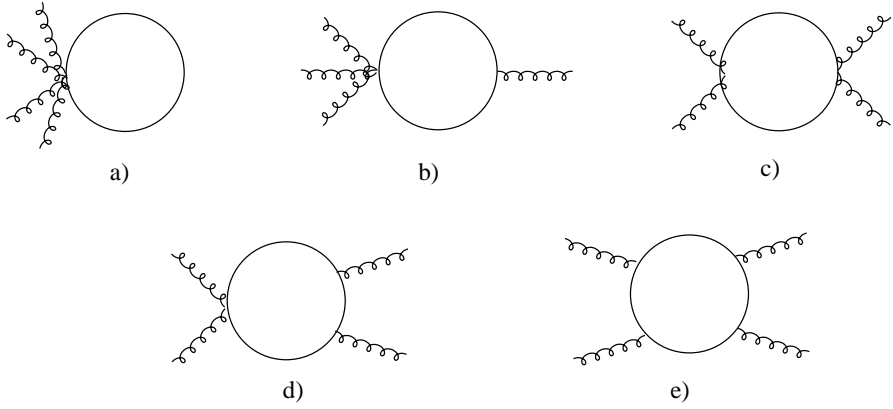}
\caption{One loop diagrams that contribute to the
effective Lagrangian ${\cal L}_4$ in Eq.~(\ref{L-phonon}). Full lines are fermionic fields. Wavy
lines are  gauge fields.}
\label{fig-4points}
\end{figure}

The Lagrangian describing the interaction of four NG bosons is formally given by
\be
{\cal L}_4 = -i\left. \left({\rm Tr} [S \Gamma] -\frac{1}2 {\rm Tr} [S \Gamma S \Gamma] +\frac{1}3 {\rm Tr} [S \Gamma S
\Gamma S \Gamma]+\frac{1}4 {\rm Tr} [S \Gamma S
\Gamma S \Gamma S \Gamma]\right)\right|_{A^4}\,,
\ee
and corresponds to the sum of the diagrams reported in Fig.~\ref{fig-4points}. All these diagrams, with the exception
of the diagram~\ref{fig-4points}e), originate from the non-local vertices $\Gamma_2$, $\Gamma_3$ and $\Gamma_4$, and
therefore can
give contributions to the coefficients of the terms with $A_0^2 \bf A^2$ and $\bf A^4$. The diagram in
Fig.~\ref{fig-4points}e) contributes to  the coefficient of the term proportional to $A_0^4$.

The contribution of the diagram in Fig.~\ref{fig-4points}a) is given by
{\small
\bea
{\rm Tr} [S \Gamma]\Big|_{A^4} &=&\dl P^{\mu\nu}A_{\mu}A_\nu (V \cdot A) \left\{ (V \cdot A) \left[\frac{- \tilde V\cdot
\ell(\tilde V\cdot \ell+2 \mu)}{D  L_0^2}\right. \right. \nonumber \\ &+&\left. Z\frac{(V\cdot \ell - 2 \mu)^2+(\tilde
V\cdot \ell + 2 \mu)^2}{D  L_0^3} \right] \nonumber\\  &+ &\left.  \tilde V \cdot A \left[  \frac{2 Z - \Delta^2 }{D  L_0^2}- 2 Z \frac{(V\cdot \ell - 2
\mu)(\tilde V\cdot \ell + 2 \mu)}{D  L_0^3} \right]+(V \to \tilde V)\right\},
\eea}
where $Z=(D+2 \Delta^2-2\mu\tilde V\cdot \ell )$.

The diagram in Fig.\ref{fig-4points}b)  gives
{\small
\bea
 & &\dl P^{\mu\nu}A_{\mu}A_\nu \left\{ (V \cdot A)^2 \left[  
 2 \Delta^2 \frac{(V\cdot \ell)^2+3(\tilde V\cdot \ell)^2-3\mu V\cdot \ell + 5\mu \tilde V\cdot \ell+ 4 \mu^2 }{D^2 
L_0^2}  \right] \right.\nonumber\\  &+&   2 \tilde V \cdot A V \cdot A  \left[ \frac{-(D+2\Delta^2)^2+\mu(V\cdot
\ell (\tilde V\cdot \ell)^2-\Delta^2(V\cdot \ell-3 \tilde V\cdot \ell )) -4  (\tilde V\cdot \ell)^2 \mu^2}{D^2  L_0^2}
\right] \nonumber\\ &+&\left.(V \to \tilde V)\right\},
\eea}
while the diagram in Fig.\ref{fig-4points}c) gives
{\small
\bea
 & &\!\!\!\!\! \dl P^{\mu\nu}A_{\mu}A_\nu P^{\alpha\beta}A_{\alpha}A_\beta \left[ \frac{D+2 \Delta^2+ 4 \mu^2 (\tilde V
\cdot \ell)^2-4 \Delta^2 \mu (2 \tilde V \cdot \ell+\mu)}{D^2  L_0^2} \right. \nonumber \\ &+& \left. \frac{V\cdot \ell(6 \Delta^2 \tilde V \cdot \ell - 4\mu (\tilde V
\cdot \ell)^2+4 \Delta^2 \mu )}{D^2  L_0^2}+ (V \to \tilde V)\right] \,.
\eea}
Both these diagrams are determined by evaluating
\be {\rm Tr} [S \Gamma S \Gamma]\Big|_{A^4}\,. \ee
The diagram in  Fig.\ref{fig-4points}d) gives
{\small
\bea
 {\rm Tr} [S \Gamma S \Gamma S \Gamma]\Big|_{A^4} &=&\dl P^{\mu\nu}A_{\mu}A_\nu \left\{ (V \cdot A)^2 \left[   \frac{
3(\tilde V\cdot \ell)^3( V\cdot \ell-2\mu)}{D^3  L_0} \right.\right. \nonumber\\ &+& \left. \Delta^2\frac{2 (V \cdot \ell)^2+7 (\tilde V \cdot \ell)^2-2 \mu
V\cdot \ell + 4  \mu \tilde V\cdot \ell   }{D^3  L_0} \right] \nonumber\\  &-&  \tilde V \cdot A V
\cdot A \left[  \frac{9 D + 12 \Delta^2 + 4 \mu (V\cdot \ell -2 \tilde V\cdot \ell) }{D  L_0^3} \right] \nonumber\\&+&\left.(V \to \tilde
V)\right\}.
\eea}
Finally, the diagram in Fig.\ref{fig-4points}e) gives
{\small
\bea
{\rm Tr} [S \Gamma S \Gamma S \Gamma S \Gamma]\Big|_{A^4} &=&\dl \frac{1}{D^4} \left[
(V \cdot A)^4 (\tilde V \cdot \ell)^4 - 3 (V \cdot A)^3 \tilde V \cdot A 
\Delta^2 (\tilde V \cdot \ell)^2  \right.\nonumber\\ & -&   V \cdot A(\tilde V \cdot A)^3 \Delta^2  (V\cdot
\ell)^2    +  (V \cdot A)^2(\tilde V \cdot A)^2(2 D + 3 \Delta^2) \nonumber\\ &+& \left.(V \to \tilde V)
 \right]\,. \eea}
After a long but straightforward  calculation we find that at the leading order in $\mu$ 
\be
{\cal L}_4 = \frac{3 }{4\pi^2}A_0^4 + \frac{3 }{4\pi^2}{\bf A}^4 - \frac{3}{2\pi^2}A_0^2
{\bf A}^2 \,.
\ee
In this case we restricted our calculation, for simplicity,  to evaluate the leading terms in $\mu$.

\section{Higgs field and interaction terms}\label{sec-higgs}
The fluctuations in the absolute value of the condensate are described by the Higgs field $\rho$ defined in Eq.~(\ref{def-delta}). 
The effective Lagrangian for this field, including the interaction terms with the NG bosons, can be determined 
by means of the same strategy employed in the previous section.

\begin{figure}[htbp!]
\centering
\includegraphics[scale=0.5]{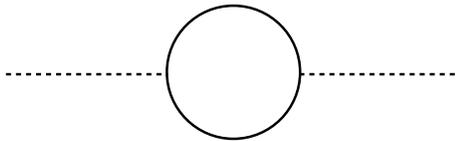}
\caption{Diagrams that contribute to the one loop self-energy of the Higgs field.  Full
lines are fermionic fields. Dashed  lines correspond to the Higgs field, $\rho$.}
\label{higgs-self}
\end{figure}

From the $\Gamma$ expansion in Eq.~(\ref{effective1}) we obtain the self-energy diagram in Fig.~\ref{higgs-self}
which gives for vanishing external momentum of the Higgs field  
\begin{equation}\label{rho-rho}
{\cal L}_{\rho\rho} =  - \frac{6}{\pi^2} \mu^2  \rho(x)^2 \,,
\end{equation}
which is the mass term for the $\rho$ field. However, the actual mass of the Higgs  is not proportional to the
chemical potential: as we shall show below  one needs a wave function renormalization in order to put the Lagrangian in
the canonical form. We notice that the $\rho$ field does not carry color and flavor indices. In order to obtain Eq.~(\ref{rho-rho}) (and  all the expressions below),  color and flavor indices have been properly contracted and the resulting Lagrangian  is expressed in terms of colorless fields. 

\begin{figure}[htbp!]
\centering
\includegraphics[scale=0.5]{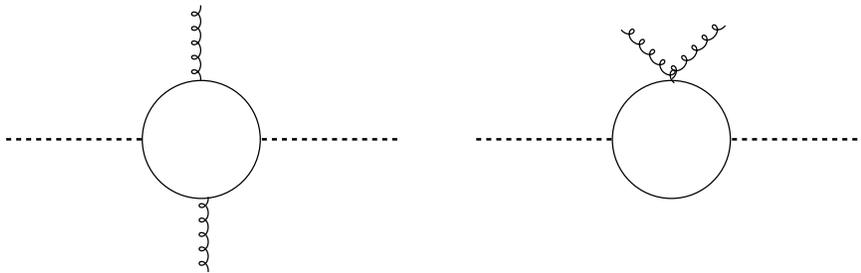}
\caption{Interaction of the Higgs field with two NG bosons at the one loop level.  Full
lines correspond to  fermionic fields, dashed  lines correspond to the Higgs field and wavy lines correspond to the
gauge field $A^\mu=\partial^\mu \phi$.}
\label{higgs-phonon}
\end{figure}

Considering  the fluctuation of the chemical potential given by $\partial_0 \phi$, the Lagrangian above turns into
\begin{equation}\label{rhorho}
{\cal L}_{\rho\rho} =  - \frac{6}{\pi^2}\tilde \mu^2  
\rho(x)^2=- \frac{6}{\pi^2}(\mu - \partial_0
\phi)^2\rho(x)^2 \,,
\end{equation}
which automatically gives part of the interaction of the Higgs field with the NG bosons. The remaining interaction
terms of two Higgs field with two NG bosons correspond to a coupling   $(\partial_i
\phi)^2\rho(x)^2$. These interaction terms, as well as the interaction terms in Eq.~(\ref{rhorho}), can be obtained from
the diagrams in Fig.~\ref{higgs-phonon}. 

Evaluating these diagrams one obtains that 
\be\label{rho2-phi2}
{\cal L}_{\rho\rho \phi\phi} =  -\frac{6}{\pi^2}\left[(\mu - \partial_0
\phi)^2
 -\frac{1}{3}(\partial_{i} \phi)^2 \right]\rho(x)^2 \,,
\ee
which describes the interaction terms among the  Higgs fields
and two NG bosons   at the leading order in $\mu$. Notice that the term $\rho
\partial_\mu \phi \partial^\mu \phi$ describing the
interaction between one Higgs field and two NG bosons is missing. This is basically due
to the fact that one cannot have a term like $\mu^2 \rho$  from the loop
expansion.
The fact that in our case  terms linear in the $\rho$ field
are missing,  leads to an interesting effect.
Let us consider a Lagrangian with terms up to  quadratic order in the NG bosons and Higgs
fields. Since the Lagrangian is quadratic in the massive Higgs field  one can integrate them out  from
the theory obtaining the actual low energy Lagrangian of the system. In the CFL phase this does
not lead to a modification of the effective Lagrangian of the NG bosons. In particular the velocity of the NG bosons
at the leading order in $\mu$ is  $c_s=\sqrt\frac{1}{3}$. This is rather different from what
happens in the non-relativistic case~\cite{Schakel:2009}, where it was shown
that integrating out the Higgs mode one obtains the modification of the
speed of sound first determined in~\cite{Marini:1998}.

In order to clarify this point, we compare our Lagrangian with the
quadratic non-relativistic Lagrangian. Schematically the
result of ~\cite{Schakel:2009} can be written as   
\be\label{lagrangianNR}
{\cal L}_{N.R.}(\rho,Y) = \frac{A}{2} \rho^2 + B \rho\, Y + \frac{C}{2}\, Y^2
+ D\, Y \,,
\ee
where $A, B, C, D$ are some coefficients and $Y = \partial_0 \phi +
\frac{(\nabla \phi)^2}{2m}$, with $m$ the mass of the non-relativistic
fermions. The speed of the NG bosons is given by 
\be\label{speed1}
c_s^2 = -\frac{D}{m C}\,. 
\ee
Integrating out the  $\rho$ field,  the effective Lagrangian for the
$\phi$ field turns out to be
\be
{\cal L}_{N.R.}(\phi) = \frac{B^2 + A C}{2 A}\, Y^2 + D\, Y \,,
\ee
and the speed of sound is modified to
\be 
c_s^2 = -\frac{D A}{m (B^2 + A C)} \,.
\ee
Notice that if the coupling $B$ vanishes, then the speed of the NG bosons remains the same one has in
Eq.~(\ref{speed1}).

In our case the LO Lagrangian quadratic in the $\rho$ field is given by Eq.~(\ref{rho2-phi2})
and there is no term that couples one $\rho$ field with two NG bosons, {\it i.e.} expanding
the Lagrangian one finds that the analogous of the coefficient $B$ in Eq.~(\ref{lagrangianNR}) is missing. 
Therefore, integrating out the $\rho$ field does not change the effective Lagrangian for the NG bosons.

From the diagrams in Fig.~\ref{higgs-phonon} we can determine the kinetic terms of the effective Lagrangian of the
Higgs filed. Considering soft momenta of the  Higgs field, $p<<\Delta$, and expanding up to the order
$(p/\Delta)^2$, we obtain 
\be
{\cal L}_{K}(\rho) = \frac{1}{2}\frac{3 \mu^2}{4 \pi^2}\frac{1}{\Delta^{2}} 
\left[(\partial_0 \rho)^2 - \frac{1}{3} ( \partial_i \rho)^2\right] -
\frac{1}{2} \frac{12 \mu^2}{\pi^2} \rho^2 \,.
\ee
By a wave function renormalization we can cast the above expression into  canonical form and we obtain that the mass of
the Higgs field is given by
\be\label{mass-rho}
m_\rho = 4 \Delta \,,
\ee
which means that the mass of the Higgs is twice the fermionic excitation energy.
This is analogous to the result obtained in the chiral sector, where the masses of the mesons turn to be equal to twice the effective  mass of the quarks, see~\cite{revNJL}.   

From the expression of the mass of the Higgs, it seems unlikely that an NG boson could excite the
$\rho$ mode. The reason is that the scale of the field $\phi$ is $T$ and in
compact stars $T \ll \Delta$. Even if thermal NG bosons in general cannot excite the
Higgs, for vortex-NG boson interaction  one has to properly take
into account the fact that as one moves inside a  vortex, the actual value of the mass of the Higgs field should decrease. The reason is that  as one moves inside a vortex,  the value of the condensate becomes smaller and smaller and  the mass of the $\rho$ field should decrease accordingly.  Therefore at a certain point it should happen that  $m_\rho < T$ and it  may become possible for an NG boson to excite the $\rho$ mode. However, a vortex is a modulation of the $\rho$ field itself, and therefore the discussion of the interaction between NG bosons and Higgs field in a vortex is subtle. We postpone to future work the analysis of this situation.

We now try to a give a general expression of  the interaction terms between two NG bosons and the Higgs fields. Let us
first neglect space variations of the $\phi$ field. Then,
expanding the effective action one has terms like
\be
{\cal L}(\rho,\partial_0 \phi) = \sum_{n \ge 2}
c_n \frac{(\mu-\partial_0\phi)^2  \rho^n}{\Delta^{n-2}}\,,
\ee
where $c_n$ are some dimensionless coefficients. This expression is simply due to the fact
that any term in the effective Lagrangian is multiplied by $\tilde\mu^2$,
which comes from the phase space integration, while the denominator comes from dimensional analysis. 

Now we want to derive the expression of the terms with space derivatives of the NG bosons.
We know from Eq.~(\ref{rho2-phi2}), that for $n=2$,  one has to replace 
$ (\mu-\partial_0\phi)^2$ with $ (\mu-\partial_0\phi)^2
-1/3({\bf \nabla}\phi)^2$, where the coefficient $1/3$ is precisely the square of the speed of sound. 
In other words the correct metric for the propagation of NG bosons is the {\it acoustic metric},
$g_{\mu\nu} = {\rm diag}(1,-1/3,-1/3,-1/3)$. Then, we introduce the four vector $X^\mu =
(\mu - \partial_0 \phi, \nabla \phi)$ and  our guess is that the Lagrangian  at the leading order in
$\mu$ is given by
\be\label{general-interaction}
{\cal L}(\rho,X^\mu) = \sum_{n\ge 2} c_n \frac{X^\mu
X^\nu g_{\mu\nu}\rho^n}{\Delta^{n-2}} \,.
\ee
The reasoning is that any perturbation propagating in the medium feels the presence of the background which induces the
acoustic metric $g_{\mu\nu}$~\cite{Unruh, Mannarelli:2008jq}.  We shall investigate in more detail this guess in future
work, however it was already shown in ~\cite{ Mannarelli:2008jq} that the free Lagrangian of  NG bosons in the CFL phase
can be written employing the acoustic metric. Therefore Eq.~(\ref{general-interaction}) seems to be an educated guess.

\begin{figure}[htbp!]
\centering
\includegraphics[scale=0.5]{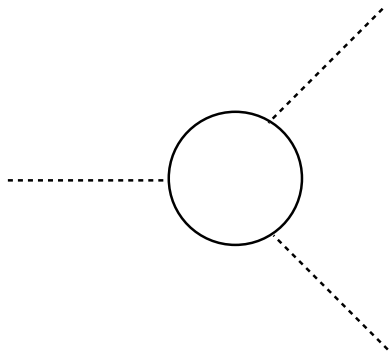}
\caption{Interaction of three Higgs fields.  Full
lines correspond to  fermionic fields, dashed  lines correspond to the Higgs fields.}
\label{higgs3}
\end{figure}

If the expression above is correct it allows to readily determine the interaction between any number of Higgs fields with
 two NG bosons in a straightforward way.
As an example we can determine the interaction between three Higgs field 
and two NG bosons. We first evaluate the  diagram in Fig.~\ref{higgs3}, which
gives the interaction among three Higgs field
\be
{\cal L}_{\rho\rho\rho} = - 
\frac{2 \mu^2}{ \pi^2 \Delta} 
\rho^3 \,,
\ee
then, upon replacing $\mu^2 \to X^\mu X^\nu g_{\mu\nu}$ we obtain the LO interaction Lagrangian
\be
{\cal L}_{\rho\rho\rho\phi\phi} = - 
\frac{2 X^\mu X^\nu g_{\mu\nu}}{ \pi^2 \Delta} \rho^3 \,.
\ee
It is not clear whether this reasoning can be extended in order to include subleading corrections of the order
$\Delta/\mu$.  In principle one would expect that the leading effect should be to perturb the metric $g_{\mu\nu}$, but further investigation in this direction is needed.

\section{Conclusions}\label{sec-conclusion}
The low energy properties of cold and dense color flavor locked quark matter are determined by the NG bosons associated
with the breaking of the $U(1)_B$ symmetry. At the very low temperatures expected in compact stars, the  NG bosons
probably give the leading contribution to the transport properties of the system.
Therefore the detailed knowledge of their effective Lagrangian is important to precisely determine the transport
coefficients.
We have determined the effective Lagrangian for this field starting from a
microscopic theory. As high energy theory we have considered the Nambu-Jona Lasinio model with a local four-Fermi
interaction with the quantum numbers of one gluon exchange. Then we have gauged the $U(1)_B$ symmetry introducing a
fictitious gauge field. Finally we have  integrated out the fermionic degrees of freedom by means of the HDET.
 
We have confirmed the results of Ref.~\cite{Son:2002zn}, which were based on symmetry arguments, and extended including
next to leading terms of order  $(\Delta/\mu)^2$. These corrections are relevant because are related with
the breaking of the conformal symmetry of the system and can be significant large for matter at non-asymptotic
densities. For this reason, the present work  paves the
way for a more detailed calculation of the transport properties of CFL quark matter, including the scale breaking
effects. 

We have also determined the interaction of the NG bosons with the Higgs mode, {\it i.e.} with the collective mode
associated with fluctuations of $|\Delta|$. These interactions are relevant in the calculation of the interaction of
the NG bosons with vortices. Indeed, vortices can be described as space modulation of $|\Delta|$. A preliminary study of
the mutual friction force in the CFL phase has been done in Ref.~\cite{Mannarelli:2008je}. However, in that calculation
of the mutual friction force only the elastic scattering of NG bosons on vortices has been taken into account. Since we
have determined the full low energy Lagrangian, we are now in a position to evaluate in more detail the interaction of 
NG bosons with  vortices, including non-elastic scattering, however we postpone the analysis of vortex-phonon interaction to future work. Indeed the treatment of this interaction is non-trivial, moreover  superfluid vortices which wind the $U(1)_B$ are topologically stable  but according to the result of the Ginzburg-Landau analysis of Ref.~\cite{Eto:2009kg} they are dynamically unstable.

\ack
We thank H.~Abuki, M.~Alford, C.~Manuel and M.~Nitta for comments and suggestions.
This work has been supported in part by the
INFN-MICINN grant with reference number FPA2008-03918E.  The work of RA has been supported in part by the
 U.\,S.\, Department of Energy, Office of Nuclear Physics, contract
 no.~DE-AC02-06CH11357. The work of MM has
been supported by the Centro Nacional de F\'isica de
Part\'iculas, Astropart\'iculas y Nuclear (CPAN) and by the
Ministerio de Educaci\'on y Ciencia (MEC) under grant
FPA2007-66665 and 2009SGR502. The work of MR has been supported by JSPS under contract number P09028.

\appendix
\section{Interaction vertices}\label{appendix-vertices}
In the HDET fermions  interact with   gauge field by the following vertices. The minimal coupling is due to the vertex
\be
\Gamma_1= \left(\begin{array}{cc} -V\cdot A & 0 \\0 & \tilde V\cdot A\end{array}\right) \,.
\ee
Non minimal couplings are due to the expansion of the non-local interaction in Eq.~(\ref{nonlocal}). Expanding in the number of gauge fields one has
the term with two gauge fields:
\be
\Gamma_{2}=\frac{P^{\mu\nu} A_\mu A_\nu}{L_0} \left(\begin{array}{cc} -2 \mu + V\cdot \ell 
 & -\Delta \\-\Delta & 2 \mu + \tilde V\cdot \ell \end{array}\right)\,,
\ee
three gauge fields: 
\be
\Gamma_{3}= -\frac{\Gamma_2}{L_0}[V\cdot A (2\mu + \tilde V \cdot \ell)+\tilde V\cdot A (2\mu - V \cdot \ell)] - \frac{\Gamma_1}{L_0} P^{\mu\nu} A_\mu A_\nu\,,
\ee
and finally the interaction with four gauge fields:
\be
\Gamma_{4}= \frac{\Gamma_{3}}{L_0} [V\cdot A (2\mu + \tilde V \cdot \ell)+\tilde V\cdot A (2\mu - V \cdot \ell)]\,.
\ee
In the present article we do not consider interactions with more than four gauge fields.

\section{Integrals} \label{appendix-integrals}
In this appendix we evaluate some integrals used in the evaluation of the effective Lagrangians for the NG boson and for the Higgs field. For the integrals in $\ell_0$ and $\ell_\parallel$ we define
\be
\int d^2 \ell \equiv \int_{-\mu}^{+\mu} d \ell_\parallel \int_{-\infty}^{+\infty} d \ell_0 \,.
\ee
We have that
\be \label{Int1}
  \int d^2 \ell  \frac{1}{D(\ell)^{n+1}} = (-1)^{n+1} \frac{i \pi}{n \Delta^{2 n}} \qquad {\rm for}\, n\geq 1 \ee
\be\label{Int2}
 \int d^2 \ell  \frac{(V\cdot \ell)^2}{D(\ell)^{n}} =\int d^2 \ell \frac{(\tilde V\cdot \ell)^2}{D(\ell)^{n}} = - i \pi
\,\delta_{n1}\,, \ee
where  $D(\ell) = V\cdot \ell \tilde V \cdot \ell - \Delta^2 + i \epsilon$.  We also need the integrals
\be
\int d^2 \ell  \frac{1}{L_0(\ell)} = - i \pi \log\left({\frac{\mu+\sqrt{\mu^2+\Delta^2}}{3 \mu+\sqrt{9\mu^2+\Delta^2}}}\right) \,,
\ee
\be
\int d^2 \ell  \frac{1}{L_0(\ell)^2} =  i \pi \frac{ \mu}{2\Delta^2} \left(\frac{1}{\sqrt{\mu^2+\Delta^2}}-\frac{3}{\sqrt{9\mu^2+\Delta^2}} \right) \,,
\ee
\be
\int d^2 \ell  \frac{1}{L_0(\ell)^3} =  -i \pi \frac{ \mu}{8\Delta^2} \left(\frac{2 \mu^2+3 \Delta^2}{(\mu^2+\Delta^2)^{3/2}}-\frac{9(6 \mu^2+\Delta^2)}{(9\mu^2+\Delta^2)^{3/2}} \right) \,,
\ee
where
\be
L_0 = (2 \mu + \tilde V \cdot \ell)(-2 \mu + V \cdot \ell) -\Delta^2 -i \epsilon \,.
\ee

\subsection{Angular integrals}
Considering a general vector $A^\mu$, for the integrals involving terms of order $A$ we have that
\be
H_0 =\int \frac{d {\bf v}}{4 \pi} V\cdot A = A_0\,.
\ee
At the order $A^2$ we have 
\bea
H_1 &=&\int \frac{d {\bf v}}{4 \pi} P^{\mu\nu}A_\mu A_\nu = -\frac{2}{3} {\bf A}^2 \,,\\
H_2 &=&\int \frac{d {\bf v}}{4 \pi} (V\cdot A)^2 = A_0^2+\frac{1}{3} {\bf A}^2\,,\\
H_3 &=&\int \frac{d {\bf v}}{4 \pi} V\cdot A \tilde V\cdot A = A_0^2-\frac{1}{3} {\bf A}^2\,.
\eea
At the order $A^3$ we have 
\bea
H_4 &=&\int \frac{d {\bf v}}{4 \pi} P^{\mu\nu}A_\mu A_\nu  V\cdot A = -\frac{2}{3} A_0{\bf A}^2 \,,\\
H_5 &=&\int \frac{d {\bf v}}{4 \pi} (V\cdot A)^3 = A_0^3 + A_0{\bf A}^2 \,,\\
H_6 &=&\int \frac{d {\bf v}}{4 \pi} (V\cdot A)(\tilde V\cdot A)^2 = A_0^3 -\frac{1}{3} A_0{\bf A}^2 \,.
\eea
At the order $A^4$ we have 
\bea
H_7 &=&\int \frac{d {\bf v}}{4 \pi} (V\cdot A)^4 = A_0^4 +2 A_0^2{\bf A}^2 + \frac{1}5 {\bf A}^4 \,,\\
H_8 &=&\int \frac{d {\bf v}}{4 \pi} (V\cdot A)(\tilde V\cdot A)^3 = A_0^4 - \frac{1}5 A_0^2 {\bf A}^2 \,,\\
H_9 &=&\int \frac{d {\bf v}}{4 \pi} (V\cdot A)^2(\tilde V\cdot A)^2 = A_0^4 - \frac{2}3 A_0^2 {\bf A}^2 + \frac{1}5 {\bf
A}^4\,,\\
H_{10} &=&\int \frac{d {\bf v}}{4 \pi} P^{\mu\nu}A_\mu A_\nu (\tilde V\cdot A)^2 = - \frac{2}3 A_0^2
{\bf A}^2 - \frac{2}{15} {\bf A}^4 \,,\\
H_{11} &=&\int \frac{d {\bf v}}{4 \pi} P^{\mu\nu}A_\mu A_\nu P^{\rho\sigma}A_\rho A_\sigma =  \frac{8}{15} {\bf A}^4 \,,\\
H_{12} &=&\int \frac{d {\bf v}}{4 \pi} P^{\mu\nu}A_\mu A_\nu \tilde V\cdot A  V\cdot A = - \frac{2}3 A_0^2
{\bf A}^2 + \frac{2}{15} {\bf A}^4 \,.
\eea


\begin{thebibliography}{99}
\bibitem{footnote}
Here $D$ is not the covariant derivative. Although this notation may
generate confusion we prefer to use it, as done in the review~\cite{Nardulli:2002review}.
\end{thebibliography}
\end{document}